# Controllable Freezing Transparency for Water Ice on Scalable Graphene Films on Copper

Bernhard Fickl,[1] Teresa M. Seifried,[1] Erwin Rait,[1] Jakob Genser,[2] Thomas Wicht,[1] Jani Kotakoski,[3] Günther Rupprechter,[1] Alois Lugstein,[2] Dengsong Zhang,[4,*] Christian Dipolt,[5] Hinrich Grothe,[1,*] Dominik Eder,[1,*] Bernhard C. Bayer[1,*]

[1]Institute of Materials Chemistry, Technische Universität Wien (TU Wien), Getreidemarkt 9, 1060 Vienna, Austria

[2]Institute of Solid State Electronics, Technische Universität Wien (TU Wien), Gußhausstraße 25-25a, 1040 Vienna, Austria

[3]Faculty of Physics, University of Vienna, Boltzmanngasse 5, 1090 Vienna, Austria

[4]International Joint Laboratory of Catalytic Chemistry, State Key Laboratory of Advanced Special Steel, Innovation Institute of Carbon Neutrality, Research Center of Nanoscience and Technology, Department of Chemistry, College of Sciences, Shanghai University, Shanghai 200444, China

[5]Rübig GmbH & Co KG, Schafwiesenstraße 56, 4600 Wels, Austria

Corresponding authors: bernhard.bayer-skoff@tuwien.ac.at (Bernhard C. Bayer), dominik.eder@tuwien.ac.at (Dominik Eder), hinrich.grothe@tuwien.ac.at (Hinrich Grothe), dszhang@shu.edu.cn (Dengsong Zhang)

**Abstract**

Control of water ice formation on surfaces is of key technological and economic importance, but the fundamental understanding of ice nucleation and growth mechanisms and the design of surfaces for controlling water freezing behaviour remain incomplete. Graphene is a two-dimensional (2D) material that has been extensively studied for its peculiar wetting properties with liquid water incl. a heavily debated wetting transparency. Furthermore, graphene is the parent structure of soot particles that are heavily implicated as nuclei in atmospheric ice formation and consequently graphene is often used as a model surface for computational ice nucleation studies. Despite this, to date experimental reports on ice formation on scalable graphene films remain missing. Towards filling this gap, we here report on the water freezing behaviour on scalably grown chemical vapour deposited (CVD) graphene films on application-relevant polycrystalline copper (Cu). We find that as-grown CVD graphene on Cu can be (as we term it) "freezing transparent" i.e. the graphene's presence does not change the freezing temperature curves of liquid water to solid ice on Cu in our measurements. Such "freezing transparency" has to date not been considered. We also show that chemical functionalization of the graphene films can result in controllable changes to the freezing behaviour to lower/higher temperatures and that also the observed freezing transparency can be lifted via functionalization. Our work thereby introduces the concept of freezing transparency of graphene on a metal support and also introduces scalable CVD graphene/Cu as an ultimately thin platform towards control of ice nucleation behaviour on a technologically highly relevant metal.

**Introduction**

Environmental ice formation from freezing of water on materials can critically impact on their operation performance, safety and running cost in many application fields.[1] For instance, ice build-up can result in structural vulnerability in large metallic structures such as overhead power line cables made from uncoated steel/Cu/Al threading, degraded energy efficiency in appliances by blocking of metallic heat exchangers in, e.g., refrigerators, or functional failure in control and lift surfaces in aerospace structures. Therefore, solutions to control ice nucleation on materials are highly sought after.[1]

To date, the control of ice formation is often an active process where materials are either heated, sprayed with anti-icing chemicals or mechanically de-iced.[1] These measures always come at a cost of energy and time, which is why a great deal of research is currently focused on creating surface treatments or extraneous coatings with the ability to *passively* control ice nucleation.[1] In this context, the desired passive control of ice nucleation on the one hand often includes "anti-icing" capabilities i.e. lowering of the temperature of ice formation on a given surface below its operation conditions. On the other hand, sometimes however also controlled "icing" at pre-determined, operationally safe locations can be desired.

The formation of ice on a material is however a highly multifaceted and as of yet not sufficiently understood process. For heterogeneous ice nucleation it is intricately linked with a material's surface energies, nano-morphologies and wettabilities.[1] Most current solutions to passive "anti-icing" surfaces include fabrication of nm/μm hierarchically structured superhydrophobic surfaces incl., e.g., slippery liquid infused porous surfaces (SLIPS).[1] Such surfaces are however hard to produce at scale, prone to damage by wear and can under certain conditions even detrimentally increase ice nucleation (e.g., droplet impingement vs. water condensation "frosting").[1] Therefore new materials/surfaces/coatings that can control ice build-up remain highly sought after. Likewise insights into the mechanistic processes of water freezing, heterogeneous ice nucleation and ice growth and how they are linked to given surface properties are still highly desired.

A multi-functional class of novel materials that has been under intense investigation over the past years are atomically-thin, two-dimensional (2D) materials such as graphene, a monolayer of $sp^2$-bonded carbon. In particular, the peculiar liquid water wetting behaviour on graphene incl. existence of a highly debated "wetting transparency" of graphene coated onto metals has recently been extensively studied and discussed.[2–9] Wetting transparency is defined as the case when the presence of graphene does not change a surface's wetting behaviour (mostly measured via contact angle) compared to the bare underlying support.[2–9] The occurrence of wetting transparency is commonly suggested to be linked to monolayer graphene's atomic thinness by which the surface properties of the underlying substrate emanate through the in comparison chemically inert graphene monolayer, thus still determining the surface's overall properties irrespective of graphene's presence.[2–13]

Despite this huge interest in graphene wetting, astonishingly practically no experimental work has been done to date on water freezing and ice nucleation on scalable graphene films on metals.[14,15] This is even more so a curious gap in literature since a sizeable number of computational studies have used graphene layers as model systems for computational ice nucleation investigation.[16–21] The lack of experimental work on ice nucleation on graphene films is furthermore surprising, since atmospheric soot composed of 2D nanocarbons, which are in first approximation nanoscopic graphene fragments, are well known nuclei for heterogeneous atmospheric ice formation, with strong (and often conflictingly observed) dependencies of ice nucleation properties on the soot's/nanographene fragments' defect levels/chemical functionalizations.[22–31]

In turn, from an application perspective, the related question arises if (functionalized) graphene on metals could be advantageously utilized to control the heterogeneous ice nucleation behaviour on metal surfaces? Such use of graphene would have high application relevance: Industrially scalable chemical vapour deposition (CVD) of graphene on a large variety of metals has been developed[32] and thus CVD graphene as a potential ice-control coating platform could be scalably employed on a range of application-wise important metals.

Despite this, the fundamental water freezing and ice nucleation properties of scalable graphene films have not been experimentally assessed yet and investigation of the exciting potentiality of scalable CVD graphene on technologically important metals as an ice nucleation control coating also remains underexplored.

Only two prior reports investigated changes in ice nucleation from graphene coatings, albeit both grown under non-scalable ultra-high-vacuum (UHV) conditions and not on a scalable metal, but on prohibitively expensive Ru/sapphire and Ir/sapphire single-crystal models.[14,15] From the literature on graphene's wetting transparency it is well known that there exists a key substrate-dependence on graphene's wetting properties.[6,24] This suggests that ice nucleation studies conducted on UHV single-crystal models may have limited applicability when it comes to making predictions about graphene-covered metal surfaces at a scalable level.

Towards filling these critical gaps in the literature, both in terms of experimental study of fundamentals of water ice nucleation on graphene as well as assessing scalable (functionalized) graphene as a potentially ice-control coating, we here report on the water freezing behaviour on scalably grown CVD graphene on application-relevant polycrystalline Cu. The employed Cu supports are hereby not only the most widely used support to produce CVD graphene[32,33] but Cu is also a highly important metal with respect to desired ice control due to Cu's widespread use in, e.g. overland power line threads or appliance heat exchangers.

We show in this report that non-treated, as-grown CVD graphene on Cu can be (as we term it) "freezing transparent" i.e. the graphene's presence does not change the freezing temperature curves of water to ice on Cu in our measurements. Such "freezing transparency" has to date not been reported, and thus also not been considered in the many computational studies that used graphene as model system for ice nucleation surfaces. Furthermore, we investigate how functionalizations to the CVD graphene (incl. -oxygen-containing, -F and -polymethylmethacrylat (-PMMA) via plasma and liquid phase treatments) can result in controllable changes of water freezing curves to lower/higher temperatures and how also the freezing transparency can be lifted. We also explore extrinsic factors necessary for observation of this freezing transparency such as storage time of our samples in ambient conditions and thus linked adventitious hydrocarbon adsorption levels. Our work thereby not only introduces the concept of freezing transparency of graphene on a metal based on first experimental observation, but also introduces scalable CVD graphene as an ultimately thin platform towards control of ice nucleation on a technologically relevant metal.

**Results and Discussion**

For our ice nucleation measurements we prepare a homogeneously covering, high-quality polycrystalline monolayer graphene film on Cu foils (25 μm thickness) by CVD (graphene/Cu).[34] For comparison we also produce bare reference Cu foil samples (Cu) with the same Cu microstructure by annealing under the same conditions as used in CVD but without the hydrocarbon exposure/graphene growth step. This similarity in Cu microstructure between CVD graphene/Cu and bare reference Cu samples is essential to allow us to attribute changes in ice nucleation temperatures solely to the presence of the CVD graphene (i.e. differences arising from Cu microstructure are thereby excluded). Microscopic and spectroscopic characterisation of the graphene/Cu and bare Cu samples is shown in Supporting Figure S1. Additionally, we investigate the effect of several functionalization treatments to the graphene/Cu stacks (and the bare Cu references). First is exposure to an air plasma (2 s), resulting in physical damage to the graphene and covalent bonding of oxygen-containing groups to the graphene defects from subsequent air exposure (Supporting Figure S2). Second is exposure to an $SF_6$ plasma (2 s), resulting in damage of the graphene and covalent formation of fluorographene (also known as "2D Teflon", Supporting Figure S3).[35] Third is functionalization of the graphene with polymeric PMMA particles via deposition and subsequent removal of a drop cast PMMA layer on the samples, which is known to result in persistent PMMA nanoparticle contamination of CVD graphene (Supporting Figure S4).[36] Samples are stored in ambient air after fabrication. The state of samples is investigated by optical microscopy, Raman spectroscopy and X-ray photoelectron spectroscopy (XPS) in parallel to ice nucleation measurements. Unless otherwise stated, ice nucleation measurements are always performed after ~24 h after the last fabrication step in order to ensure a comparable level of inevitable adventitious hydrocarbon contamination adsorption from ambient air storage accumulated on the samples.[5,6] This is key as prior work on wetting transparency of water on graphene has shown that different adventitious carbon contamination levels can significantly alter graphene's wetting behaviour, making comparison for non-ambient-exposure-time controlled samples difficult.[2–6,8,9] For selected samples in our study also the time evolution of freezing behaviour as a function of storage time in ambient (2 h to 1 month) and corresponding hydrocarbon contamination is assessed in order to disentangle the effect of the inevitable hydrocarbon adsorption. Further details on sample preparation and characterisation can be found in the Supporting Information.

Ice nucleation measurements[37] on these various samples are performed in an optical cryo-microscopy setup (schematic in Figure 1a),[38] consisting of a cryo-cell containing a Peltier element (Quick-cool QC-31-1.4-3.7M) that can cool down to –40 °C through thermoelectric cooling and is temperature controlled via feedback from a K-type thermocouple mounted directly on the sample stage. The cryo-cell has a glass window and is mounted directly on an optical light microscope stage, so that the freezing of individual water droplets can be observed as a function of temperature in the optical microscope. The freezing stage is housed inside an air tight cell at atmospheric pressure, which is purged with dry $N_2$ gas before ice nucleation measurements. This results in a low humidity atmosphere inside the chamber and thus suppresses secondary water droplet formation from condensation during cooling runs.[14] As the water reservoir for freezing, each sample is sprayed with ultra-clean water (MiliQ$^{fi}$ 18.2 MΩ·cm) at room temperature before being inserted into the freezing cell, leading to a distribution of water droplets on the sample surface (example in first frame in Figure 1b). We thermally contact the Cu foil backside to the stage with a small amount of paraffin oil to ensure thermal contact between Peltier stage and Cu samples. During the freezing experiments, the Peltier element is cooled at a steady rate of 10 K/min. Concurrently, optical microscope images of the water droplets are recorded at a rate of 20 images/s during cooling incl. a temperature stamp in each recorded image. The freezing of an individual droplet can readily be detected in the optical microscope image sequences by a change in optical appearance whereby the water droplet changes contrast upon freezing (Figure 1b).[14,39] The freezing event of a given water droplet is therefore here assigned to this change in optical appearance of a given water droplet, in line with prior literature.[14,39] We note that freezing of small water droplets is a fast process occurring over a timescale of fractions of a second.[14,39] Since our cooling rate of 10 K/min is on a much slower timescale, we can assign the temperature at which this freezing event is observed as the freezing temperature of this given water droplet. An exemplary freezing series incl. temperature stamps is shown in Figure 1b. As the contrast change upon freezing can be quite subtle, a freezing event can be further accentuated by employing image difference calculations (Figure 1c). Using the optical microscope image sequences, we thereby manually and with help of a custom-programmed image analysis algorithm (see Supporting Information) measure freezing temperatures (and diameters) for multiple (10 to >40) water droplets per individual sample from such freezing optical microscopy video series. Multiple individual samples are measured per every sample fabrication run. This data is then presented as freezing curves for each sample condition (e.g., Figure 2), showing the number fraction (in %) of frozen water droplets (frozen droplets fraction) versus temperature. The here presented freezing curves all consist of data from repeated runs and plot interpolated, averaged curves (detailed explanation on data treatment in Supporting Information). We emphasize that to date no experimental work has studied such ensemble freezing curves for water on graphene films.[32,33] Estimated uncertainties (based on standard deviation) to the freezing curves and extracted values are presented throughout the manuscript and Supporting Information. For applications the onset temperature of freezing for a macroscopic water deposit is often important. To extract a simple comparative estimate for such onset of freezing for a given droplet ensemble on a macroscopic sample we extract from our freezing curves a $T_{10}$ value at which 10% of water droplets have frozen for a given sample. Likewise, we also extract $T_{25}$, $T_{50}$ and $T_{75}$ values at which 25%, 50% and 75% of droplets have frozen, respectively. We tabulate these temperature values for the various studied samples conditions in Table 1 and Supporting Table S1. Further details on the freezing experiments and their analysis procedure can be found in the Supporting Information.

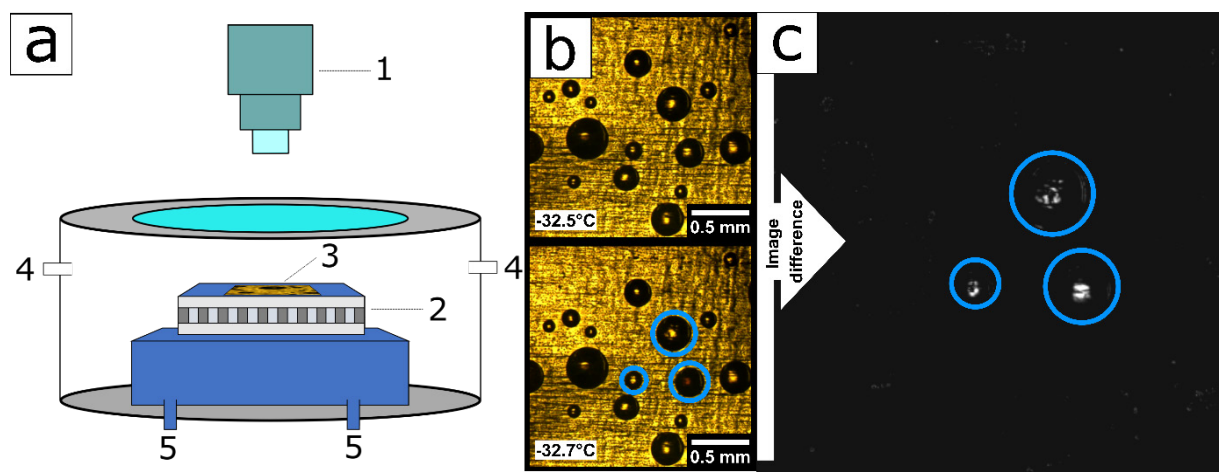

**Figure 1.** (a) Schematic illustration of cryo-microscopy setup (1: microscope objective, 2: Peltier cooling element, 3: copper foil sample, 4: in-/outlet for $N_2$ purging, 5: in-/outlet water-cooling). (b) Optical micrograph series, incl. temperature-stamps, depicting sample region before (upper frame) and after multiple exemplary freezing events of the marked (blue circles) water droplets. (c) Image difference calculations based on the data presented in (b) accentuating the freezing event of the marked water droplets for easier detection.

**Freezing on as-grown graphene/Cu.** Figure 2 compares freezing curves of the CVD graphene/Cu stack (solid black line) compared to the bare Cu reference (dashed black line). The freezing curve data shown in Figure 2 shows averaged freezing fraction curves of >20 separate runs for each sample type with 10 to 30 droplet freezing events in each run and also includes uncertainty bands (shaded bands, calculated from standard deviations). What is strikingly apparent from Figure 2 is that no significant difference in the freezing behaviour between CVD graphene/Cu stacks (solid black line) and bare Cu references (dashed black line) is observed. For both sample types, little freezing is observed before reaching –25 °C, with then a small increase in frozen droplets between –25 °C to –30 °C and then rapid freezing starting at around –30 °C and ending with all droplets frozen around –34 °C. For comparison, homogeneous ice nucleation temperature of water in the absence of a heterogeneous surface to nucleate is commonly reported at –36 °C to –38 °C.[40] The data shows that under our measurement conditions both graphene/Cu and Cu allow significant undercooling of water droplets below 0 °C before heterogeneous ice nucleation occurs. Importantly, comparing graphene/Cu and Cu, we neither observe differences in the onset temperatures of freezing, nor significant differences in the further freezing fraction evolution upon further cooling. This suggests that non-treated, as-grown CVD graphene on Cu is (as we term it) "freezing transparent" compared to bare Cu references at our measurement conditions. We here introduce the term "freezing transparency" in analogy to the prior studied "wetting transparency" of graphene[2–9] and define it as the case when the presence of the graphene on a given support does not change the freezing temperature evolution of water compared to on the bare support. The freezing transparency behaviour of the graphene on Cu in our measurements is also reflected in the calculated $T_{10}$, $T_{25}$, $T_{50}$ and $T_{75}$ values in Table 1 and which are identical within our error margin for Cu/graphene and the bare Cu reference. We note that we do not evidence a distinct difference in water droplet sizes for graphene/Cu stacks and bare Cu, as both show largely similar droplet size distributions (Supporting Figure S5). Within the respective size distributions we find a weak trend of larger droplets freezing at higher temperatures similarly for both samples (Supporting Figure S6), which is well in line with general heterogeneous ice nucleation theory, where a larger contact area from a larger water droplet is related to a larger propensity for ice nucleation.[41]

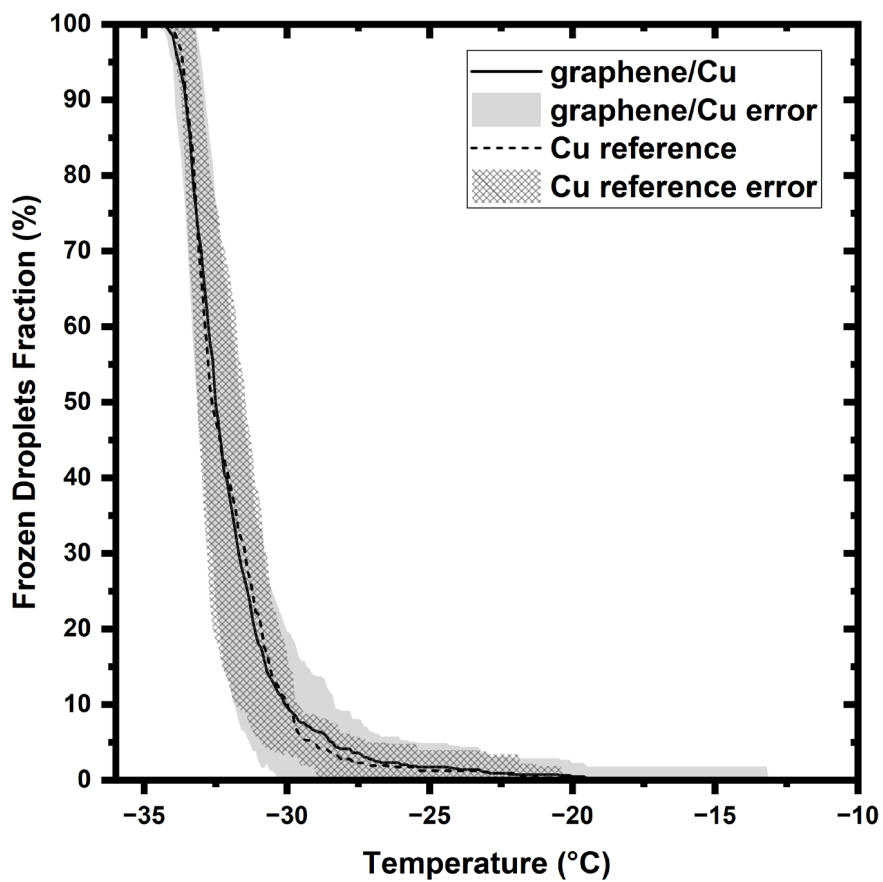

**Figure 2.** Freezing curves of non-treated, as-grown CVD graphene/Cu stacks (solid black) and bare Cu references (dashed black). Curves are averages of >20 separate runs for each sample type with 10 to 30 droplet freezing events in each run from multiple sample preparations. Shaded areas represent standard deviation bands to the freezing curves.

**Table 1.** $T_{10}$, $T_{25}$ $T_{50}$ and $T_{75}$ values and their standard deviations ($\Delta T_{xx}$) for our various samples for graphene/Cu and bare annealed Cu as fabricated and after the various functionalizations (all measured 24 h after last fabrication step).

| | $T_{10}$/°C | $\Delta T_{10}$/°C | $T_{25}$/°C | $\Delta T_{25}$/°C | $T_{50}$/°C | $\Delta T_{50}$/°C | $T_{75}$/°C | $\Delta T_{75}$/°C |
|---|---|---|---|---|---|---|---|---|
| Graphene/Cu as-grown | −30.1 | ± 2.0 | −31.6 | ± 1.1 | −32.4 | ± 0.8 | −33.0 | ± 0.7 |
| Graphene/Cu air plasma | −30.9 | ± 1.7 | −32.3 | ± 0.7 | −33.0 | ± 0.5 | −33.4 | ± 0.5 |
| Graphene/Cu PMMA | −26.4 | ± 3.2 | −28.4 | ± 3.5 | −29.3 | ± 3.5 | −32.1 | ± 0.5 |
| Graphene/Cu SF$_6$ plasma 2 s | −30.4 | ± 0.6 | −31.1 | ± 0.5 | −32.4 | ± 0.3 | −33.1 | ± 0.3 |
| Graphene/Cu SF$_6$ plasma 10 s | −15.5 | ± 2.5 | −17.6 | ± 5.2 | −20.6 | ± 6.3 | −23.8 | ± 7.9 |
| Cu annealed | −30.5 | ± 1.4 | −31.6 | ± 1.1 | −32.3 | ± 0.9 | −32.9 | ± 0.5 |
| Cu air plasma | −30.4 | ± 4.0 | −32.1 | ± 2.0 | −33.3 | ± 0.4 | −33.8 | ± 0.4 |
| Cu PMMA | −26.9 | ± 3.8 | −30.4 | ± 4.3 | −33.9 | ± 0.3 | −34.2 | ± 0.2 |
| Cu SF$_6$ plasma 2 s | −9.6 | ± 2.2 | −9.6 | ± 2.2 | −9.6 | ± 2.2 | −9.6 | ± 2.2 |
| Cu SF$_6$ plasma 10 s | −13.8 | ± 5.1 | −13.8 | ± 5.1 | −13.8 | ± 5.1 | −13.8 | ± 5.2 |

To date a freezing transparency of graphene to water ice, as observed here, has not been reported in the literature. We therefore now first discuss our here reported observation of freezing transparency of graphene for water ice in light of the existing and related literature on the wetting transparency of graphene for liquid water.[2–9]

Wetting transparency of graphene is defined as water having the same wetting behaviour (defined typically by contact angle) to a given substrate even if a monolayer of graphene is sandwiched between the water and the given substrate (i.e. analogous to our above introduced definition of freezing transparency).[2–9] Across a series of reports, graphene has been reported to be fully wetting transparent, partially wetting transparent or not wetting transparent at all.[2–9] Full wetting transparency is said to occur when the contact angle of the graphene/substrate stack is the same as of the bare substrate, partial wetting transparency when the measured contact angle is between freestanding graphene's and bare substrate's respective contact angles and no wetting transparency when the measured contact angle is the same as on freestanding graphene irrespective of its substrate. The disparities in literature regarding presence, partial or absence of wetting transparency have emerged to be related to several factors contributing to the wettability of graphene incl. not only i. graphene's intrinsic wettability but also ii. environmental factors.[2–9] In terms of environmental factors in particular a. adsorbed adventitious carbon contamination build-up and b. the type of substrate underneath the graphene have been identified to play key roles in graphene's wetting transparancy.[2–9] An emerging consensus in the literature is that graphene can be at least partially wetting transparent when i. the graphene is monolayered, ii. the graphene's interaction with its substrates is weak (i.e. the substrate does not strongly alter the graphene's electronic structure) and iii. the graphene has only low levels of adventitious carbon adsorbate build-up on top.[2–9] Cu, as used in our study as substrate, is such a weakly interacting substrate for graphene.[6,33] Consequently, for fresh, monolayered CVD graphene on Cu (partial) wetting transparency has been experimentally observed.[2,6] In contrast for strongly interacting Ni no wetting transparency of monolayered CVD graphene was observed.[6] Notably however, the observed wetting behaviour of graphene/Cu showed a significant time dependence for samples stored under ambient conditions, with over time changing presence or absence of (partial) wetting transparency over timeframes from h to years.[5,6] This has been shown to be related to adventitious carbon contamination from ambient building up on all samples over time, which with long enough storage time in ambient leads to the adventitious carbon contamination's wetting properties completely overshadowing the sample's wetting properties.[5,6]

To ensure a constant level of adventitious carbon contamination in our sample series,[5,6] all above presented measurements have therefore been acquired ~24 h after sample fabrication. To further explore the evolution of the freezing behaviour on our samples as a function of longer storage time, we present in Supporting Figure S7 freezing curves for storage in ambient until 1 month. We find that after 1 week freezing curves on Cu are similar to the ~24 h measurements, while for graphene/Cu after 1 week a shift of the freezing curve to slightly higher temperature is observed. This suggests that the freezing transparency has been lifted by the longer ambient air storage. After 1 month of storage the onset temperatures of freezing for Cu and the entire freezing curves for graphene/Cu have shifted to higher temperatures. Importantly, also after 1 month the curves for graphene/Cu and Cu do not overlap any more i.e. freezing transparency has disappeared (Supporting Figure S7). We suggest that this upshift of freezing temperatures and disappearance of freezing transparency is related to the massive adventitious carbon contamination build-up from extended ambient storage.[5,6] Importantly, the 1 month data in Supporting Figure S7 thereby also shows that the freezing transparency at shorter storage times is not only a result of adventitious carbon contamination, since such contamination would be most dominating for the 1 month sample, where freezing transparency was however lifted.

On the other hand, in Supporting Figure S8, we probe the freezing behaviour of our graphene/Cu and Cu samples after shorter storage times of only ~2 h in ambient. Based on prior literature,[5,6] we know that for these samples the adventitious carbon contamination levels will be as low as technically possible for our measurement facilities (given how fast we can reliably bring samples from fabrication to ice nucleation measurements). We find that for such samples the onset temperature for freezing and entire freezing curves are significantly shifted to higher temperatures. Also, the freezing curves for graphene/Cu and Cu do not overlap perfectly any more for the ~2 h stored samples, indicating an at least partial absence of the freezing transparency of the ~2 h stored samples.

Combined, Supporting Figure S7 and S8 therefore indicate that the freezing transparency for graphene/Cu and graphene is only observed for intermediate storage time in ambient (~24 h). We suggest that this is linked to the medium adventitious carbon contamination levels after ~24 h in ambient. We emphasize however that for massive adventitious carbon build-up (after 1 month) the freezing transparency vanishes, which in turn suggests that the observed freezing transparency is *not* just an effect of freezing on substrate-independent adventitious carbon. In fact, combining Figure 2 and Supporting Figures S7 and S8 suggests that in order to establish the here observed freezing transparency not only the graphene in the graphene/Cu vs. Cu systems is necessary but also a mediating level of adventitious carbon on the graphene. This conclusion is reminiscent of the findings in the prior wetting transparency literature,[2–9] suggesting that similar underlying mechanisms are at play in the case of freezing on graphene films.

**Freezing on functionalized graphene/Cu.** So far, we have investigated the freezing properties on as-grown graphene on Cu, which displays the here reported freezing transparency and elucidated its time dependence on inevitable adventitious carbon contamination. In order to test if the graphene's freezing behaviour on a metal can also be controlled in a deliberate fashion beyond storage time, Figure 3 compares droplet freezing curves for graphene/Cu stacks and bare Cu references that additionally underwent deliberate functionalization treatments before the ice nucleation measurements. For comparison, the non-treated, as-grown graphene/Cu and bare Cu freezing curves from Figure 2 are re-plotted as solid and dashed black lines, respectively. All measurements in Figure 3 are taken after ~24 h storage in ambient conditions after the last fabrication/treatment step to ensure a fair comparison to the data in Figure 2.

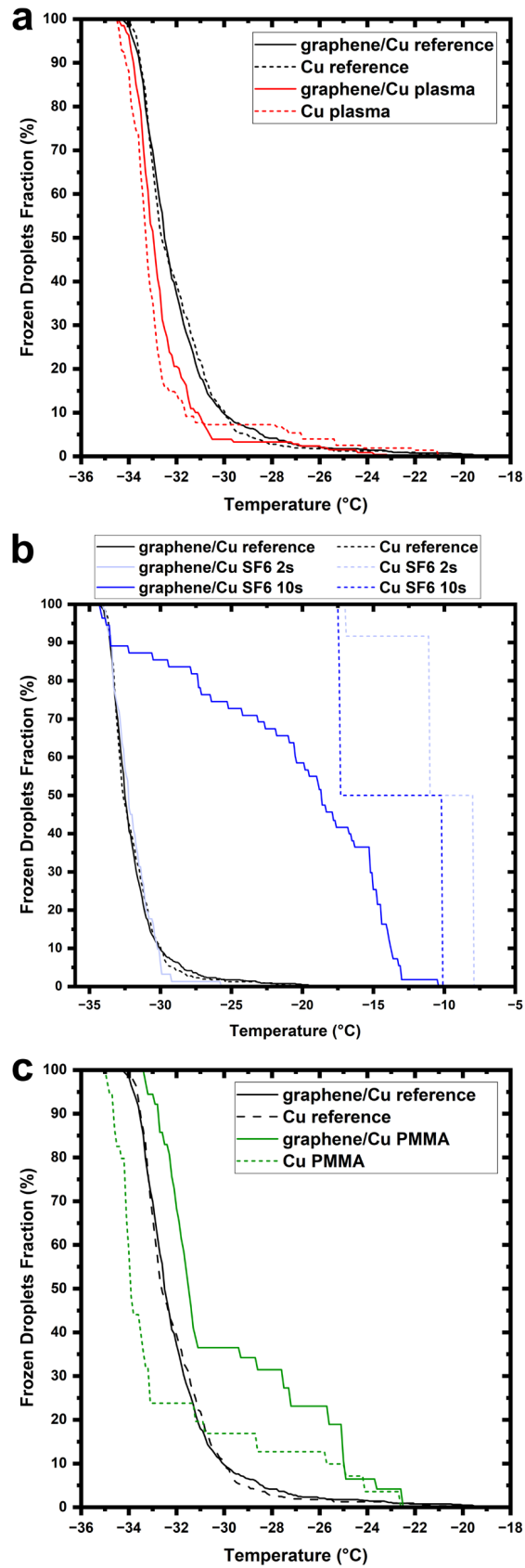

**Figure 3.** (a) Freezing curves for air plasma treated samples (x-axis scale –18 °C to –36 °C). (b) Freezing fraction curves for SF$_6$ plasma treated samples (x-axis scale –5 °C to –36 °C). (c) Freezing

curves for PMMA treated samples (x-axis scale –18 °C to –36 °C). Graphene/Cu samples are represented by solid lines and Cu samples by dashed lines, respectively. All diagrams include freezing curves of untreated, as-grown graphene/Cu (black solid) and Cu (black dashed), replotted from Figure 2. All curves are averages of several experimental freezing runs; a version of this figure with uncertainty bands included is shown in Supporting Figure S9.

The first treatment we investigate is covalent functionalization of the CVD graphene with oxygen-containing groups by air plasma treatment (2 s) (Figure 3a, red solid and dashed curves). As shown via Raman spectroscopy in Supporting Figure S2, such air plasma treatment incl. subsequent ambient air exposure results in severe damage to the graphene lattice and thus covalent defect functionalization of the graphene with oxygen-containing functional groups.[42] A similar state of samples has prior also been labelled as "graphene oxide"-like.[15] Such treatment is motivated by prior observation of freezing behaviour changes in atmospheric soot by different oxygen-containing groups.[22–31] Notably, we find in Figure 3 and Table 1 that such air plasma treated graphene/Cu (red solid curve) does show freezing at slightly lower temperatures (by around –1 K) compared to non-treated, as-grown graphene/Cu (solid black) for the same ambient air storage time. Importantly also, between air-plasma-treated graphene/Cu and air-plasma-treated Cu little difference is observed, i.e. both Cu/graphene and Cu are similarly shifted to slightly lower freezing temperatures. We note that measured water droplet size distributions are not significantly different for air-plasma-treated Cu/graphene and air-plasma-treated Cu (Supporting Figure S10 and S11) compared to their non-plasma-treated counterparts (Supporting Figure S5 and S6). This suggests that it is not a different water droplet size distribution that is indirectly responsible for the slightly lower freezing temperatures, but rather suggests that the introduced oxygen-containing groups on the air-plasma-treated surfaces chemically mediate the freezing behaviour to slightly lower temperatures. The oxygen plasma experiments show that functionalization of graphene (here with oxygen-containing groups) can downshift the freezing temperature curve on graphene to slightly lower temperatures.

The second treatment we investigate is covalent functionalization of the graphene by F i.e. formation of highly hydrophobic fluorographene or "2D Teflon".[35] This is motivated by hydrophobicity often being linked to lower freezing temperatures.[1] We fabricate this material by exposure of graphene/Cu and bare Cu references to a $SF_6$ plasma (2 s and 10 s). Supporting Figure S3 confirms that this results in strong damage to the graphene lattice and covalent functionalization of the graphene with F towards fluorographene for the 2 s $SF_6$ plasma.[35] The corresponding freezing curves in Figure 3b of the 2 s $SF_6$-plasma-treated sample shows for graphene/Cu (solid light blue curve) a surprisingly similar freezing behaviour as on non-treated, as-grown graphene/Cu (solid black curve). This result in turn shows that even with a covalent functionalization the freezing behaviour of graphene does not necessarily change. This is quite noteworthy, as in contrast the 2 s $SF_6$-plasma-treated Cu reference (dashed light blue curve in Figure 3b) shows a drastically different freezing behaviour with a very much higher freezing temperature interval already between –10 °C to –15 °C. Similarly, the 10 s $SF_6$-plasma-treated graphene/Cu (solid dark blue) and the 10 s $SF_6$-plasma-treated Cu (dashed dark blue) samples show drastically upshifted freezing behaviour. Based on XPS in Supporting Figure S3, we attribute this high temperature onset of freezing on $SF_6$-plasma-treated Cu (2 s and 10 s) and on $SF_6$-plasma-treated graphene/Cu (10 s) to formation of $CuF_2$. Notably, $CuF_2$ formation also results in a drastic change in water droplet sizes (Supporting Figure S12). The presence of comparatively still intact graphene during $SF_6$ plasma (2 s) prevents formation of $CuF_2$, while for longer $SF_6$ plasma (10 s) treatments the graphene has been strongly destroyed, making room for $CuF_2$ formation. The $SF_6$-plasma experiments show that functionalization of graphene (here for F, 2 s plasma) can also leave the freezing temperature curve unaffected compared to untreated graphene.

The third treatment that we apply to the graphene is wet chemical, non-covalent functionalization with PMMA particles that are persistently anchored onto the graphene.[36] We achieved this by drop-casting and hot plate-hardening of PMMA in anisole on the graphene/Cu and bare Cu reference samples, followed by subsequent removal of hardened PMMA in acetone/isopropanol. This is well known from prior work to lead to persistent contamination of the CVD graphene with PMMA microparticles and nanoparticles, while the graphene lattice remains structurally perfectly intact (see also our characterisation data in Supporting Figure S4).[36] In the corresponding freezing data in Figure 3c we find that for such PMMA-functionalized graphene/Cu (solid green curve) this translates to a significant fraction of the water droplets already nucleating at between –25 °C and –30 °C, which is at a significantly higher temperature than for untreated, as-grown graphene/Cu (solid black curve). This is also reflected by the higher $T_{10}$ temperature for PMMA-treated graphene/Cu compared to as-grown graphene/Cu (Table 1). We attribute this to the persistent PMMA particles to act as preferential nucleation sites for ice nucleation, as compared to the comparatively atomically smooth, chemical inert basal plane of non-treated, as-grown CVD graphene. This is also reaffirmed by the freezing data on PMMA-treated bare Cu (dashed green curve) which also shows a higher temperature onset of freezing (albeit less pronounced than for the PMMA-treated graphene/Cu). The PMMA experiments thus show that functionalization of graphene (here with PMMA particles) can also lead to a higher onset temperature of freezing compared to untreated graphene.

Combined, the results in Figure 3 indicate that the here first observed freezing behaviour of water on graphene on a metal surface (Figure 2) can be modified by functionalization treatments. While for addition of oxygen-containing groups we observe a slight down-shift of freezing temperatures (air plasma in Figure 3a), for addition of polymeric residues we observe an up-shift of freezing temperatures (PMMA in Figure 3c). The former down-shift is suggested to be related to a change of chemical interaction of water with the introduced oxygen-containing groups, while the latter up-shift is suggested to be related to polymeric particles acting a preferential nucleation centres for ice nucleation. Interestingly for functionalisation with F, the freezing behaviour on graphene is not affected while on the Cu substrate it is strongly changed ($SF_6$ plasma in Figure 3b). The observation that introduction of oxygen-containing groups (Figure 3a) can decrease freezing temperatures is in line with a prior paper[15] that reported a lower freezing temperature for oxidized graphene vs. pristine graphene on UHV Ir single crystals (note that possible freezing transparency between graphene/Ir vs. Ir was not studied in this ref. [15]). The results in Figure 3c on the effect of deliberate PMMA contamination to alter freezing behaviour via increased ice nucleation, links well with prior work on PMMA residues (and similar polymer residues) to have an significant effect on graphene wetting.[8] Also, the results in Figure 3b regarding $SF_6$ plasma (flourographene) are interesting compared to prior literature: While in our experiments we do not evidence any significant change in freezing behaviour for the fluorinated graphene on Cu, in contrast a prior study[14] reported a lower freezing temperature for flourographene vs. pristine graphene on UHV Ru single crystals (note that possible freezing transparency between graphene/Ru vs. Ru was not studied in this ref. [14]). This discrepancy suggests that substrate effects (Ru vs. Cu) could be important in the freezing behaviour of graphene, which is also in line with the importance of substrate effects in wetting behaviour of graphene.[6,8]

We also want to discuss the limitations of our here presented study. We report hitherto not observed water freezing transparency of graphene/Cu vs. Cu under our conditions. We emphasize however that presence or absence of such freezing transparency is suggested to be contingent on several factors. As already shown here, a key factor is storage time in ambient and its linked adventitious carbon contamination build-up. We further however suggest that the type of substrate will be key to the observed freezing behaviour. While here freezing transparency on weakly interacting Cu is observed, more interacting substrates such as Ni[6] or Fe as supports may result in different freezing behaviour on such supported CVD graphene. Furthermore, in comparison to wetting experiments, freezing experiments also have a much wider parameter space in terms of environmental conditions and experimental approaches.[1] While wetting experiments almost exclusively use contact angle measurements, for freezing experiments the experimental pathways to bring the water to surface and induce its freezing are much more varied: We here employ freezing of pre-supported water droplets in a dry atmosphere. Freezing experiments can however also include other water delivery pathways such as condensation freezing from a highly humid atmosphere or droplet impingement freezing on pre-cooled surfaces.[1] The type of droplet delivery and cooling pathways have been prior shown to significantly alter freezing behaviour on various surfaces incl., e.g., widely studied SLIPS.[1]

We therefore emphasize that our reported findings here are a first experimental impetus to explore the currently critically under-investigated experimental freezing behaviour on scalable graphene and other 2D materials films, incl. the here found freezing transparency, in a similar fashion as wetting behaviour of graphene and other 2D materials has been extensively studied in the recent past.[2–9,2–6,8,9] We also note that such experimental investigations will provide key feedback to all the computational work that currently uses graphene as model surfaces for fundamental investigation of ice nucleation and on the hitherto overlooked freezing transparency.

We also note that the here introduced freezing transparency concept not only links to prior work on wetting behaviour/wetting transparency of graphene, but in a wider realm also to other phenomena which rely on substrate properties to emanate through atomically thin graphene/2D materials:[13] These include, e.g., graphene-substrate-assisted growth modes of extraneous films on graphene from vapour phase techniques[10,11] as well as "remote" epitaxy in which epitaxial relations between a substrate and a deposited film are kept despite the presence of a sandwiched graphene interlayer.[12] These processes similarly rely on a phase transition of an extraneous material (here water/ice) on an atomically thin 2D material sandwiched between the extraneous material and a bulk support, as explored here for water freezing on graphene/Cu.

**Conclusions**

In summary, we have experimentally studied the water freezing behaviour on scalable CVD graphene films on application-relevant Cu. It was found that as-grown CVD graphene on Cu can be "freezing transparent", which is a term that we introduce to describe the phenomenon when freezing curves on graphene/Cu vs. bare Cu reference samples are identical i.e. the presence of graphene does not change the water freezing behaviour compared to on its bare underlying substrate. We explored the conditions in which such freezing transparency can be observed and also explored how chemical functionalization of the graphene films can result in changes to freezing evolution to lower/higher temperatures. Our work thus introduces the concept of freezing transparency of graphene on a metal based on the first experimental observation and also introduces scalable CVD graphene as an ultimately thin materials platform for control of ice nucleation and water freezing behaviour on a technologically relevant metal.

**Author Contributions**

B.C.B. conceived the idea and supervised the work. Samples were prepared by B.F. and B.C.B. Freezing measurements were done by B.F. and T.S. on a set-up developed by H.G. Freezing data was analysed by B.F., T.S. and B.C.B. with assistance from E.R. Authors J.G., T.W., J.K., G.R., A.L., D.Z., C.D. and D.E. assisted with sample preparation, sample characterisation and data interpretation. The manuscript was written by B.F. and B.C.B. with input from all authors.

**Acknowledgements**

B.C.B., D.Z. and C.D. acknowledge funding from the Austrian Research Promotion Agency (FFG) under project 879844-HARD2D. B.C.B. also acknowledges partial funding to the work from the European Research Council (ERC) under project 101088366-HighEntropy2D.

**Supporting Information**

**Supporting Methods**

**Graphene CVD**

We employed 25 μm thick Cu foils (Alfa Aesar Puratonic 99.999%) as catalysts for graphene CVD in a custom-made hot wall tube furnace at reduced pressure (base pressure ~$10^{-3}$ mbar), based a prior reported CVD recipe.[1] The Cu foils were first pre-treated at 960 °C in 2000 sccm flow of Ar with 5% $H_2$ resulting in ~14 mbar pressure in order to facilitate Cu grain growth and reduction of Cu-oxides from foil storage in ambient air. For graphene growth 50 sccm $CH_4$ were added for 7 min, increasing the total pressure to ~15 mbar. After growth, the Cu foils were left to cool in Ar/$H_2$ atmosphere by sliding the tube furnace's hot zone from the samples. This recipe results in a closed film of high-quality graphene on Cu.[1] Graphene-free reference Cu samples with the same Cu grain structure were prepared using the same processing with the exception of the $CH_4$ exposure step.

**Functionalization treatments**

**Air plasma.** Samples were air plasma treated in a commercial plasma chamber (Atto from Diener electronic GmbH & Co KG, Germany). A plasma (at 50% power level of 40 kHz 0-200 W; 13.56 MHz 0-300 W) was struck in ~0.5 mbar air with exposure of the samples to the air plasma for 2 s.

**$SF_6$ plasma.** For $SF_6$ plasma treatment a PlasmaPro 100 Cobra (OXFORD Instruments) system was used with a pressure of ~0.05 mbar, a $SF_6$ gas flux of 40 sccm and a bias of 21 V with RF = 18 W.

**PMMA.** PMMA functionalization of graphene/Cu and annealed Cu samples was done by drop casting PMMA photoresist (200K, AR-P 642.04, Allresist GmbH, Germany) onto the foil samples and curing the PMMA on a hotplate at 100°C in air. Samples were subsequently put into acetone for 2 h and rinsed with DI water and isopropanol to remove the PMMA film, which is however known to be an imperfect process, resulting in persistent decoration of the graphene with PMMA micro- and nanoparticles.[2]

Samples were stored in ambient air.

**Materials Characterisation**

**Optical Microscopy and Raman Spectroscopy.**[3,4] Optical microscopy and Raman spectroscopy were conducted using a WITec alpha 300 RSA+ system with laser wavelength 488 nm, laser power 10 mW and spot size ~2 μm and in a NT MDT Ntegra Spectra system with laser wavelength of 473 nm.

**XPS.** X-ray photoelectron spectroscopy (XPS) measurements were performed with a Specs XR50© high intensity non-monochromatic Al/Mg dual anode and an X-ray source Phoibos 100 energy analyzer (EA) with multichannel plate. The spectra were obtained at room temperature, an emission angle of 0°, a pass energy of 20 eV and using an Al anode with $K_\alpha$ radiation at 1486.6 eV. Data analysis was performed via CasaXPS. Calibration of the spectra was deemed unnecessary, as due to the high conductivity of the copper foils minimal sample charging was anticipated. All spectra were analyzed with an energy step width of 0.1 eV, though only every third data point is shown in the figures for improved clarity.

**Ice nucleation measurements**

Samples are measures ~24 h after the last process step unless otherwise stated. Samples are sprayed with MiliQ[fi] (18.2 MΩ·cm) water by hand, using a spray bottle and fine nozzle creating a mist of fine droplets. This results in reproducible droplet sizes and densities on the samples overall. The samples are mounted onto a thermoelectric cooler (TEC; i.e., a Peltier element Quick-cool QC-31-1.4-3.7M) with a drop of paraffin oil for optimal thermal contact. The cooler sits directly on a water powered heat exchanger and is enclosed in an air tight container with a built-in glass window for samples observation with an optical microscope (Figure 1a). Temperature is controlled via a K-type thermocouple mounted directly onto the cooler surface with thermally conductive adhesive. The housing of the cryo-cell[5] allows for atmospheric control via two gas connectors which are used to flush the cell volume with dry nitrogen before every experimental run, substantially decreasing humidity. During a freezing experiment the Peltier element cools the sample at a steady rate of 10 K/min to –40 °C, at which all droplets have undergone freezing. Temperature and video are recorded via LabVIEW VI (virtual instrument) and stores as Video with temperature- and timestamp, plus a separate temperature log file for data processing.

**Freezing Data Processing**

Freezing experiments are recorded as video (20 fps) with temperature- and time stamp and separate temperature log file. Temperature is recorded in ~0.15 K steps, which is therefore our limit for temperature resolution in our measurements. The video is then analysed, either manually with the help of image processing software (ImageJ/Fiji[6,7]) by converting the video to an image stack and running an image difference operation ("stack difference") before going through the stack frame by frame to extract individual freezing events, or with the help of an automated Python script. The algorithm in a first step extracts the drop location and size (diameter) from the first frame of the video. As the drops appear much darker than the copper background, only simple image processing is needed to extract this information. First, the image is converted to greyscale, blurred slightly to remove any contrast from the texture of the copper sheet and a threshold is applied to convert the image to a binary black-white image. In this state, only the drops should appear as black blobs while the copper sheet should be mostly white. To further isolate the drops, a morphological erosion operation is performed and the connected black regions that are large enough (in the number of pixels) are retained and assigned a label. From this the diameter and position is computed, assuming spherical drops and a region of interest (ROI) is drawn for each drop. At this point the user gets to check if the drop assignment was done correctly before the freezing event detection resumes. To detect freezing events the algorithm analyses the change in brightness between 2 subsequent frames. If inside one ROI, which was computed as previously described, a large enough absolute change in brightness occurs, this is detected as a freezing event. Together with the temperature log file the freezing temperature for individual freezing events is recorded. The output is then manually checked for irregular or double detection of freezing events as contrast change upon freezing may happen over a timespan of more than one frame. This raw data in form of droplet freezing events is converted to freezing fraction curves (0% to 100%) and interpolated between 0.1 K temperature steps with the condition that the freezing fraction is considered constant between freezing events. This results in freezing fraction curves that show a stepwise increase with decreasing temperature. These freezing fraction curves of different experimental runs are then averaged to give the displayed average freezing curves. A visualization of this process is shown in Supporting Figure S13. It should be noted that not all average freezing curves contain the same number of experimental runs and freezing events and therefore vary in their statistical significance which is reflected in the difference in standard deviation (error bands). $T_{10}$, $T_{25}$ $T_{50}$ and $T_{75}$ represent the temperature at which 10%, 25%, 50% and 75% of the droplets are frozen, respectively. The temperature values are extracted for each experimental run and then averaged to obtain the values in Table 1 and Supporting Table S1. The error values represent the standard deviation from the mean value.

**Supporting Figures**

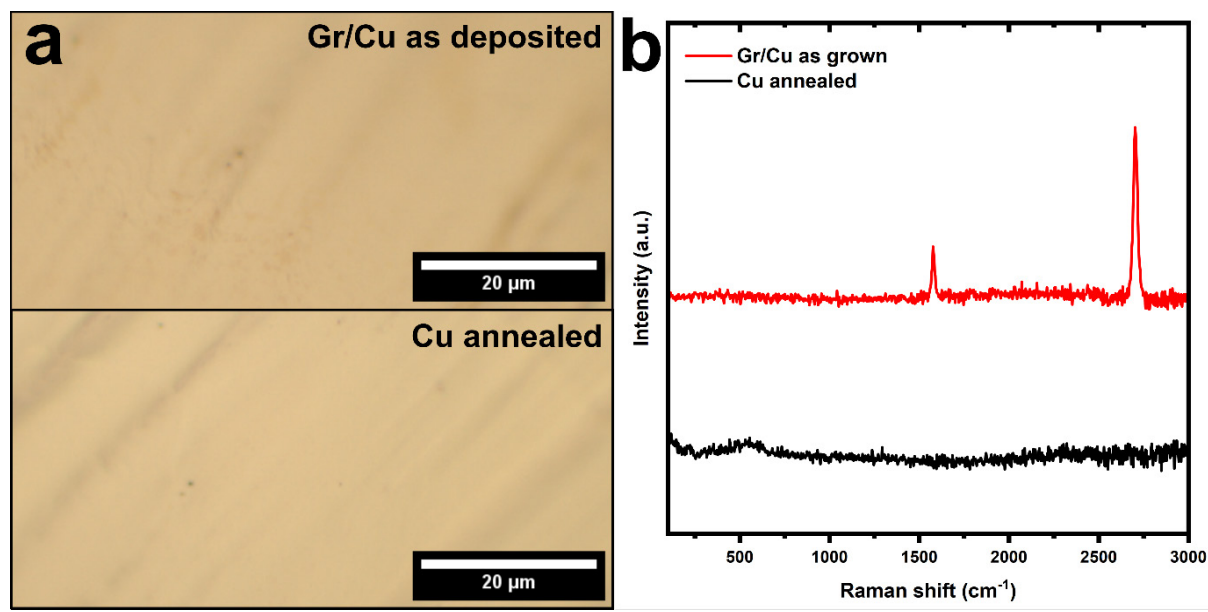

**Supporting Figure S1.** (a) Optical microscopy and (b) Raman spectroscopic characterisation of as-deposited graphene/Cu (top) and annealed bare Cu (bottom) samples. The Raman spectrum in (b) clearly confirms that our CVD graphene films on Cu are almost exclusively monolayer and of high quality.[1,3] Additional transfer experiments of the monolayer graphene films from the Cu to $SiO_2$(90 nm)/Si wafers (not shown) allow us to confirm via optical microscopy[8] that the coverage of the Cu samples with monolayered graphene is >99% areal coverage. Raman also confirms the bare annealed Cu sample to be void of significant carbon deposits (beyond adventitious hydrocarbon adsorbates).[3]

S5

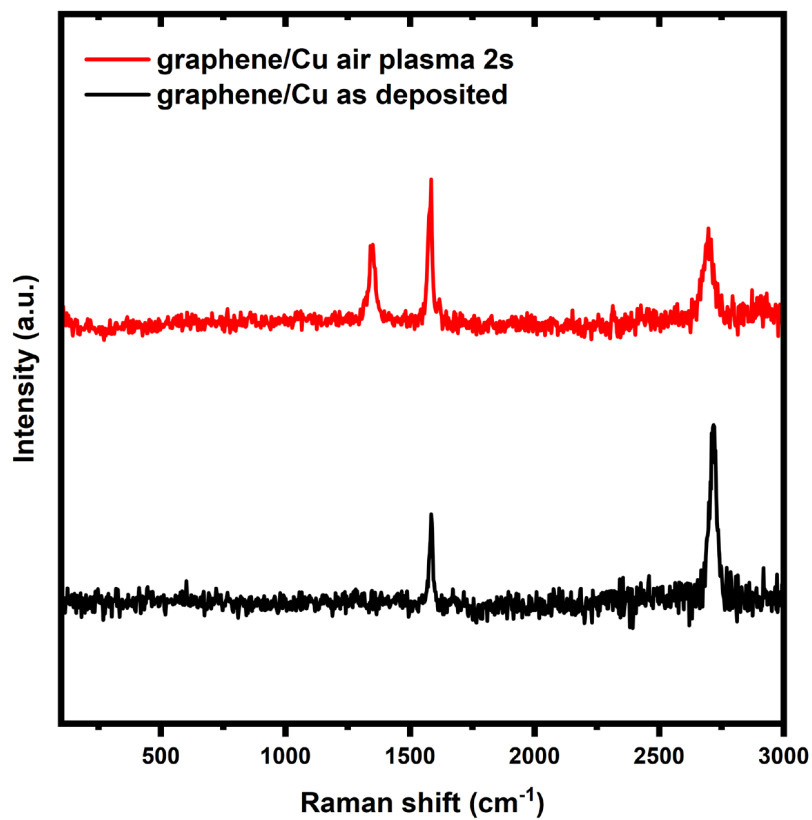

**Supporting Figure S2.** Raman spectra of graphene/Cu samples after 2 s air plasma, confirming plasma induced damage to the graphene via appearance of a pronounced defect-related D peak (~1350cm$^{-1}$).[3] These defects are known to be readily decorated with oxygen-containing groups during the air plasma and when exposed to ambient air.[9]

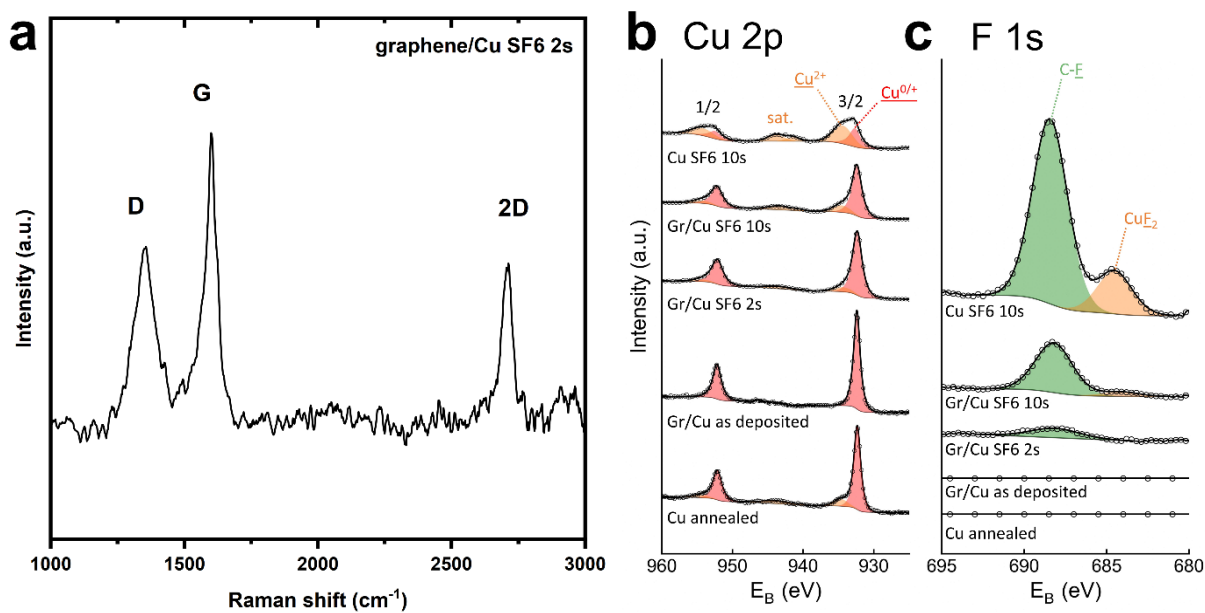

**Supporting Figure S3.** (a) Raman spectrum of graphene/Cu sample after 2 s $SF_6$ plasma treatment, confirming plasma induced damage to the graphene via appearance of a pronounced defect-related D peak (~1350cm$^{-1}$) and suppression of 2D peak (~2700cm$^{-1}$).[3] (b) Cu2p and (c) F1s X-ray photoelectron spectra of graphene/Cu ("Gr/Cu") and bare annealed Cu samples after 10 s plasma (top) and 2 s plasma (middle) compared to as deposited graphene/Cu and bare Cu samples (bottom). See discussion paragraph below for XPS and Raman data interpretation.

**$SF_6$ plasma treatment discussion.** The Raman spectrum of graphene/Cu sample after 2 s $SF_6$ plasma treatment in Supporting Figure S3a confirms plasma induced damage to the graphene via appearance of a pronounced defect-related D peak (~1350cm$^{-1}$) and suppression of 2D peak (~2700cm$^{-1}$)[3] in line with partial formation of fluorographene.[10] In Supporting Figure S3 panels (b) and (c), for all samples the Cu2p region consists of a doublet at around 932.4 eV and 952.2 eV matching with either $Cu^0$ or $Cu^+$. Indeed, the Cu LMM region (not shown) contains two peak maxima at a kinetic energy of 916.5 eV and 918.3 eV pointing to the presence of both Cu species.[11,12] While all non-$SF_6$-plasma-treated samples are void of any F signal, the $SF_6$-plasma-treatment leads to formation of an organic fluoride species ~688.5 eV (likely C-F) in the F1s region.[10,13] For the graphene/Cu samples this is related to graphene reacting with F towards flourographene. For the Cu samples adventitious hydrocarbon deposits react with the F. Moreover, $SF_6$-plasma treatment results in formation of $CuF_2$, as evident from both the

S7

presence of a metal fluoride species ~684.5 eV in the F1s region and by the $Cu^{2+}$ doublet and satellite feature.[13,14] That said, the graphene layer appears to present a barrier against the fluorination of copper as the amount of $CuF_2$ present in graphene/Cu $SF_6$ 10s is significantly lower than that of Cu $SF_6$ 10 s. In summary, the XPS data is consistent with the Raman data in showing incorporation of F into the graphene layer for $SF_6$ plasma 2 s towards formation of flourographene.[10]

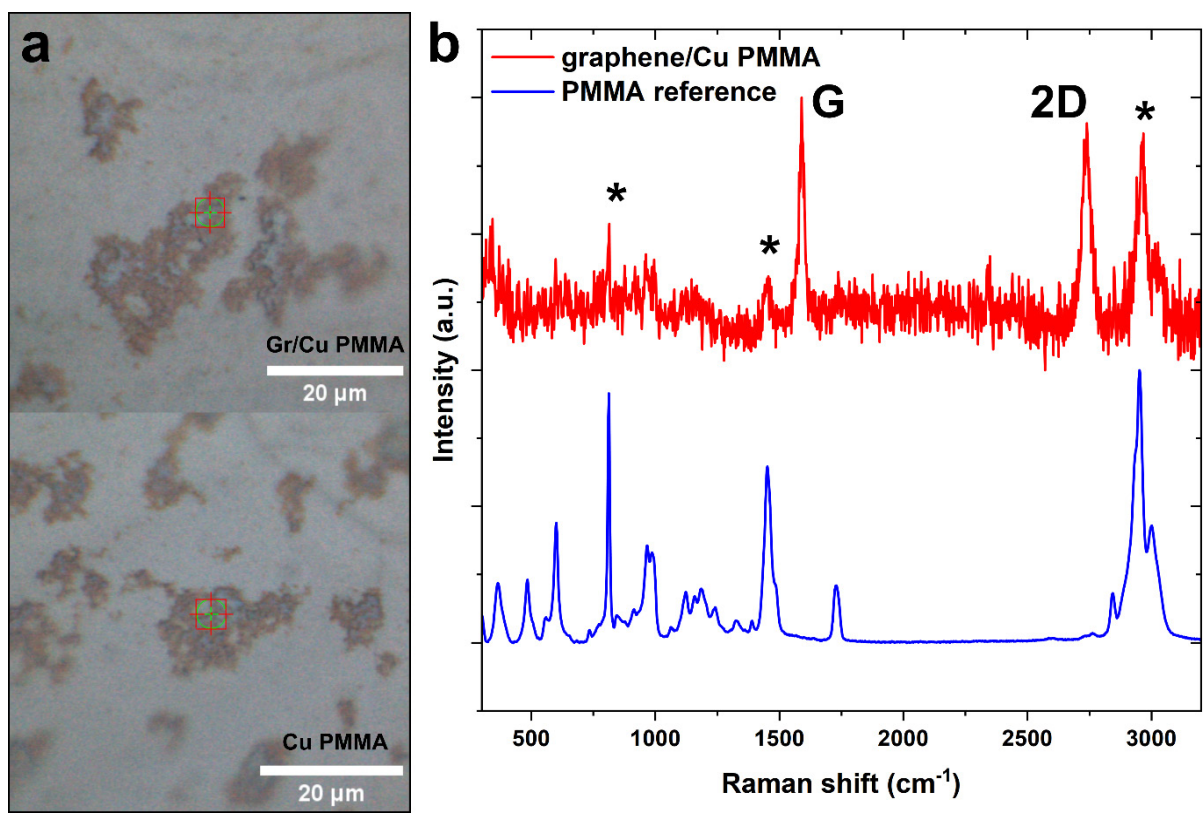

**Supporting Figure S4.** (a) Optical micrographs of graphene/Cu (top) samples decorated with PMMA functionalization via deposition and subsequent removal of a drop cast PMMA layer.[2] Large persistent PMMA deposits are visible on the surface in addition to the well-known nanoscopic residual PMMA nanoparticles from this treatment.[2] Reference bare Cu (bottom) shows a similar amount of macroscopic PMMA residue after PMMA drop cast and removal. (b) Raman spectrum of graphene/Cu (top) sample decorated with PMMA, confirming the presence of PMMA functionalization on the CVD graphene and reference Raman spectrum of PMMA film[15] (bottom). PMMA peaks visible in graphene/Cu Raman spectrum are marked with '*'. We note that the PMMA signal overlaps in the region of the D peak of graphene (~1350cm$^{-1}$), therefore complicating interpretation. It is however well known that graphene does not structurally degrade from such PMMA functionalization.[3]

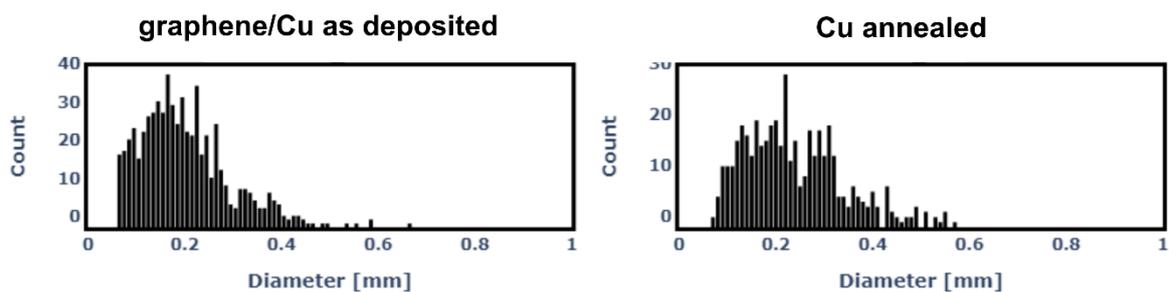

**Supporting Figure S5.** Histograms of water droplet diameters for graphene/Cu (left panel) and bare annealed Cu (right panel).

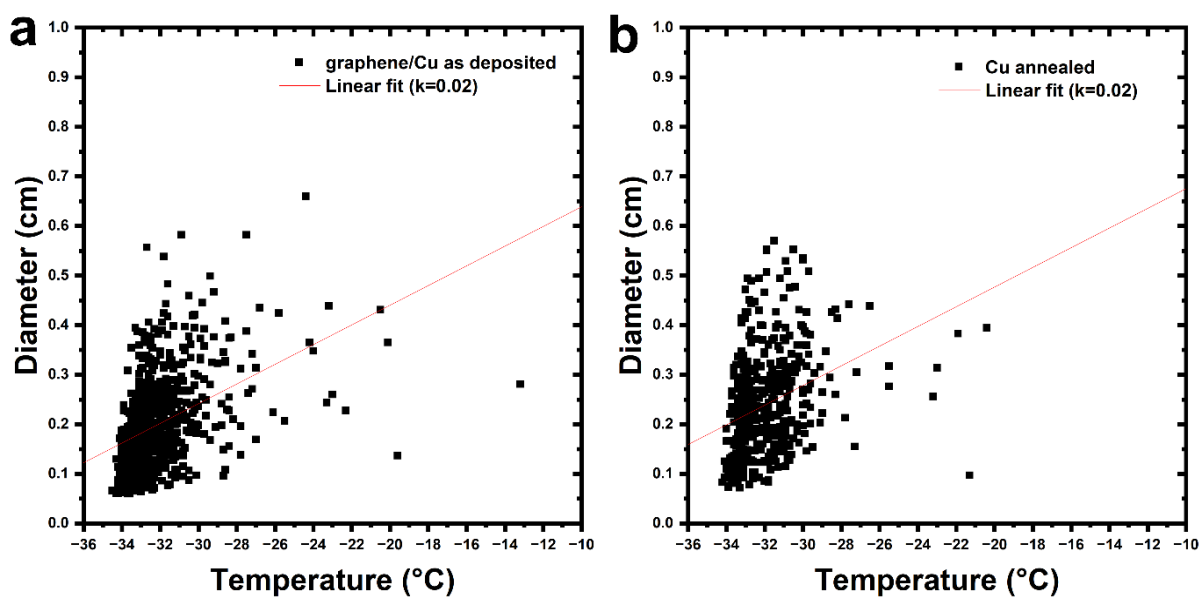

**Supporting Figure S6.** Dependence of freezing temperature on droplet diameter for as deposited graphene/Cu (right panel) and bare annealed Cu (left panel). Linear fit and slope (k) as guide for the eye.

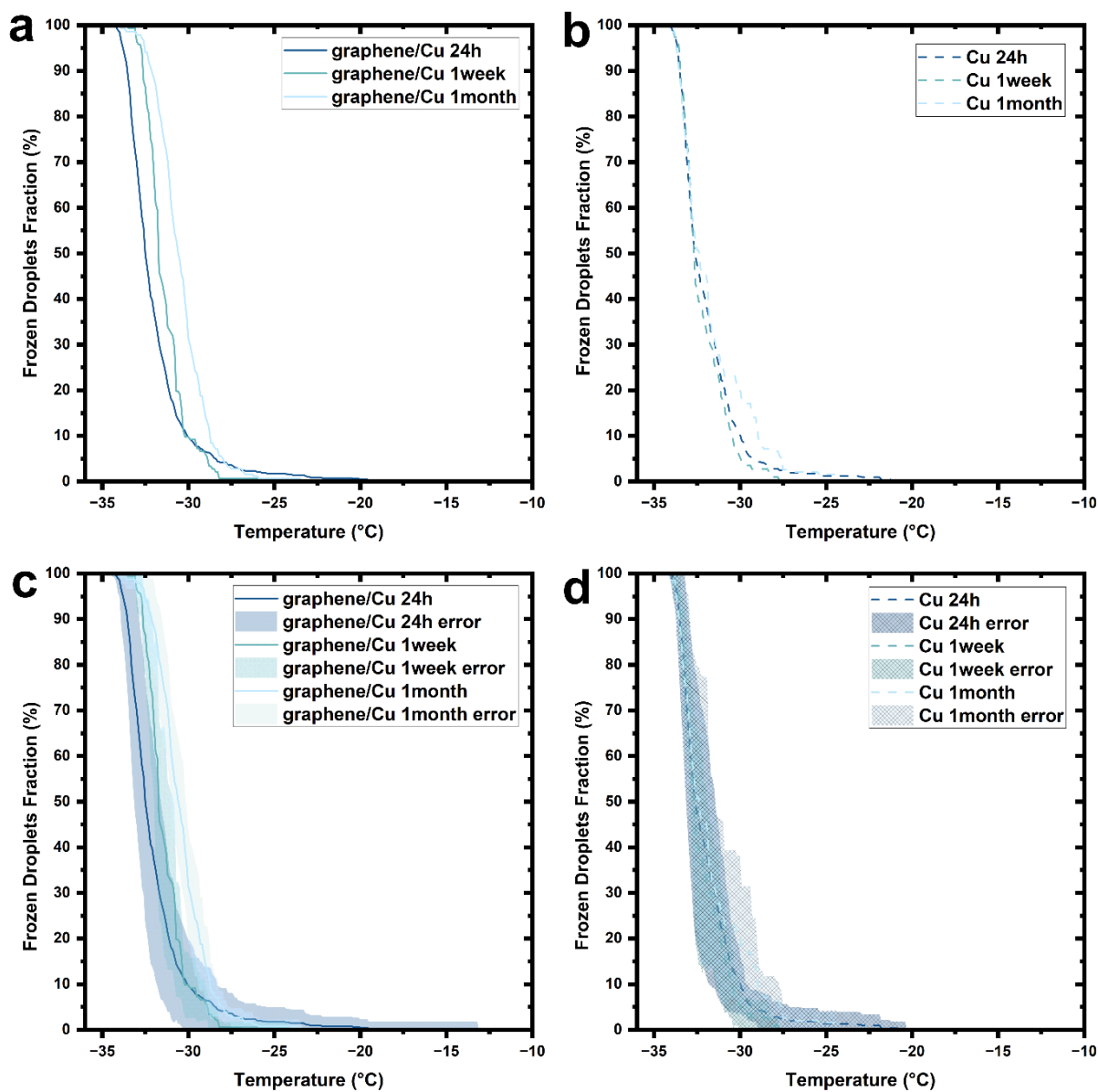

**Supporting Figure S7.** Freezing curves on (a) graphene/Cu as deposited and (b) Cu annealed as a function of storage time in ambient conditions from ~24 h to 1 month (24 h data replotted from Figure 2). Panels (c) and (d) show respective plots of (a) and (b) with error bands representing their standard deviation.

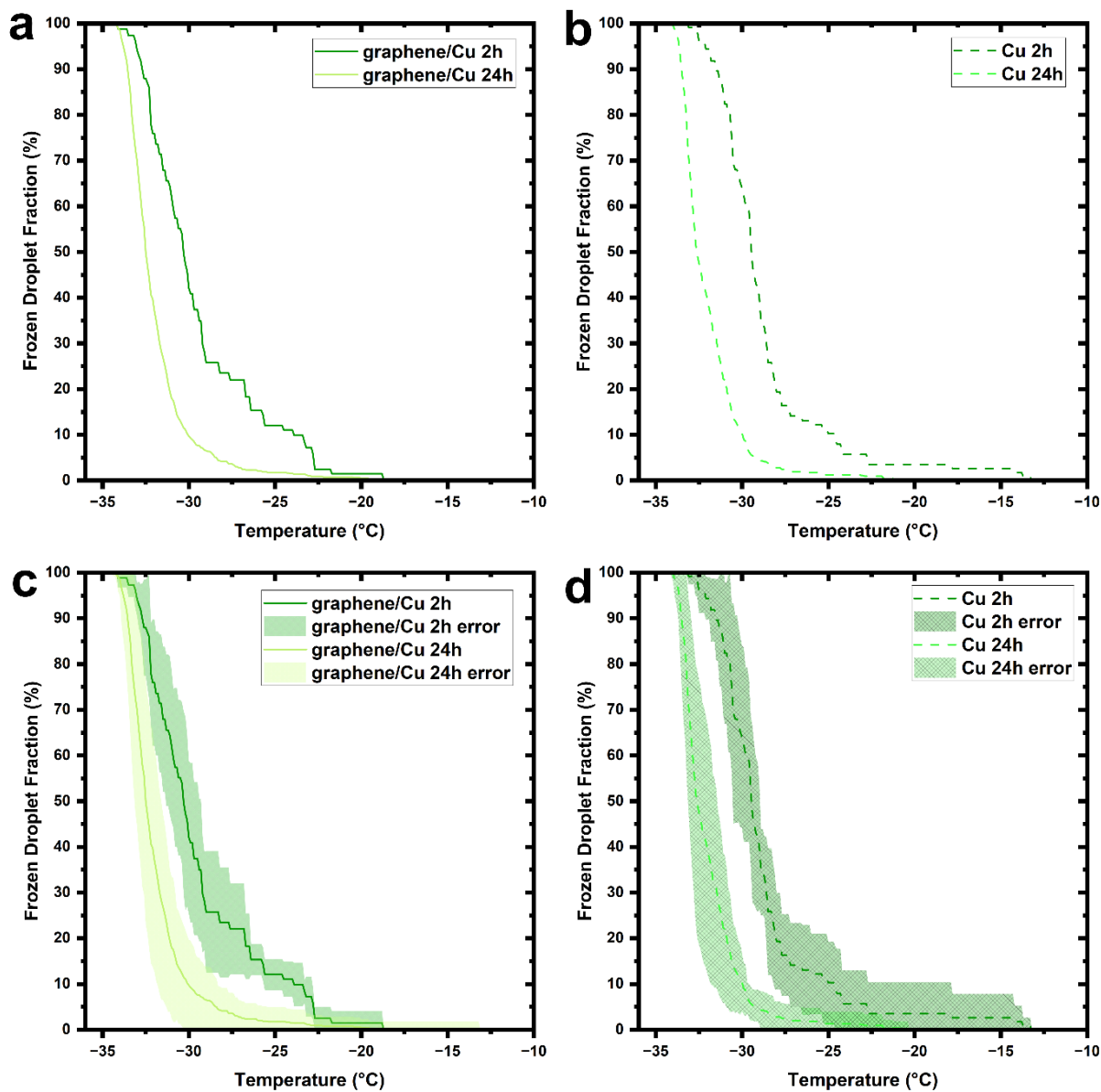

**Supporting Figure S8.** Freezing curves on (a) graphene/Cu as deposited and (b) Cu annealed as a function of storage time in ambient from ~2 h to ~24 h (24 h data replotted from Figure 2). Panels (c) and (d) show respective plots of (a) and (b) with error bands representing their standard deviation.

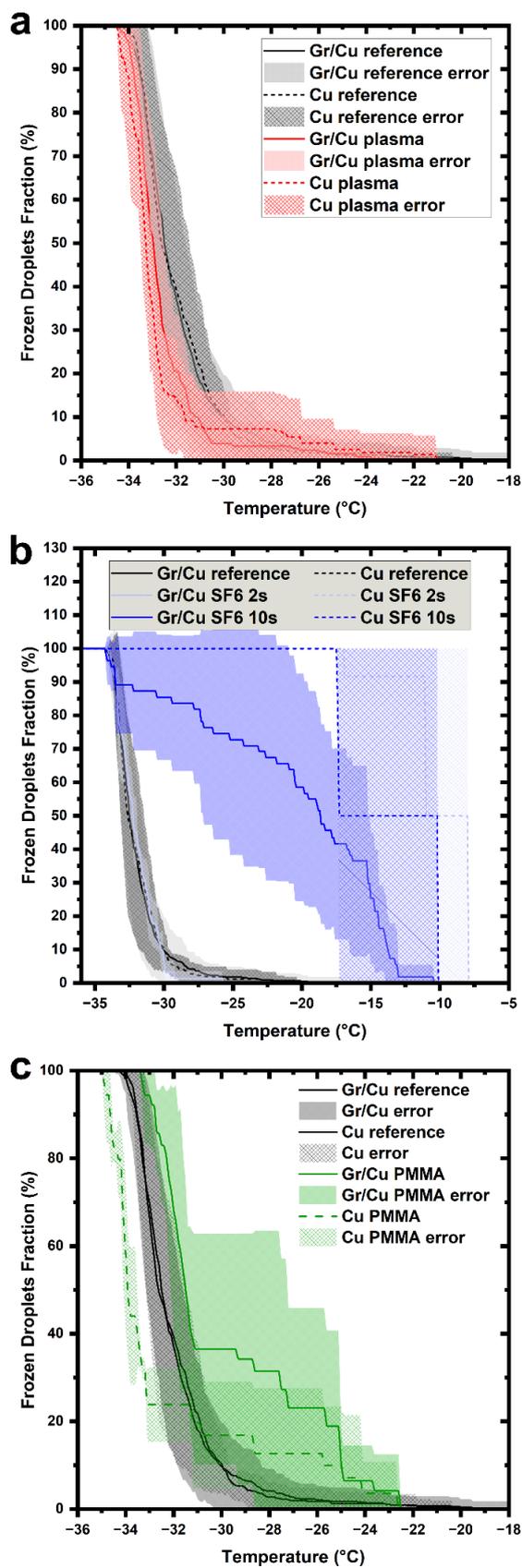

**Supporting Figure S9.** Replot from Figure 3 with uncertainty bands from standard deviations as shaded areas added. (a) Freezing curves for air plasma treated samples (x-axis scale −18 °C to −36 °C).

(b) Freezing fraction curves for SF$_6$ plasma treated samples (x-axis scale –5 °C to –36 °C). (c) Freezing curves for PMMA treated samples (x-axis scale –18 °C to –36 °C). Graphene/Cu samples are represented by solid lines and Cu samples by dashed lines, respectively. All diagrams include freezing curves of untreated, as-grown graphene/Cu (black solid) and Cu (black dashed), replotted from Figure 2. All curves are averages of several experimental freezing runs.

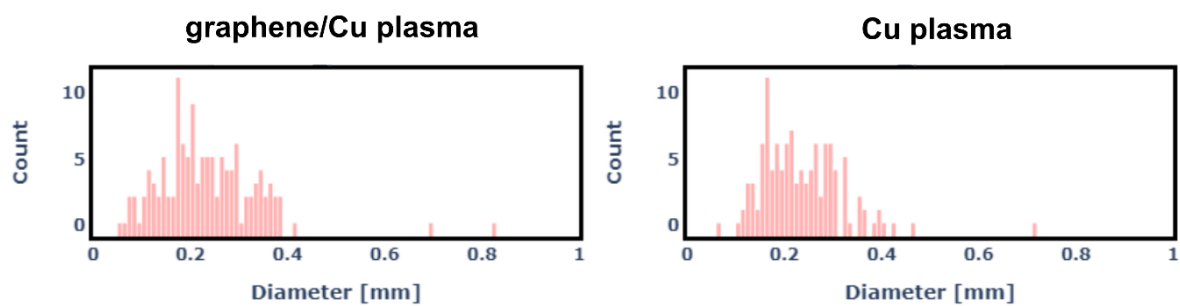

**Supporting Figure S10.** Histograms of water droplet diameters for air-plasma-treated graphene/Cu (left panel) and air-plasma-treated Cu (right panel)

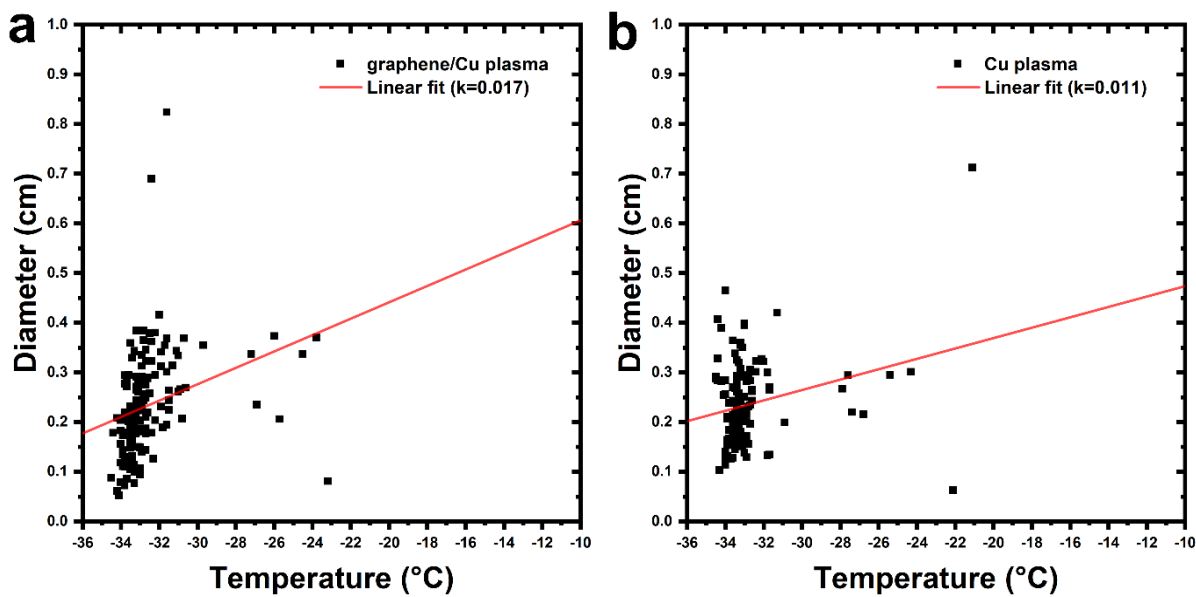

**Supporting Figure S11.** Dependence of freezing temperature on droplet diameter for air plasma treated graphene/Cu (left panel) and air plasma treated annealed Cu (right panel). Linear fit and slope (k) as guide for the eye.

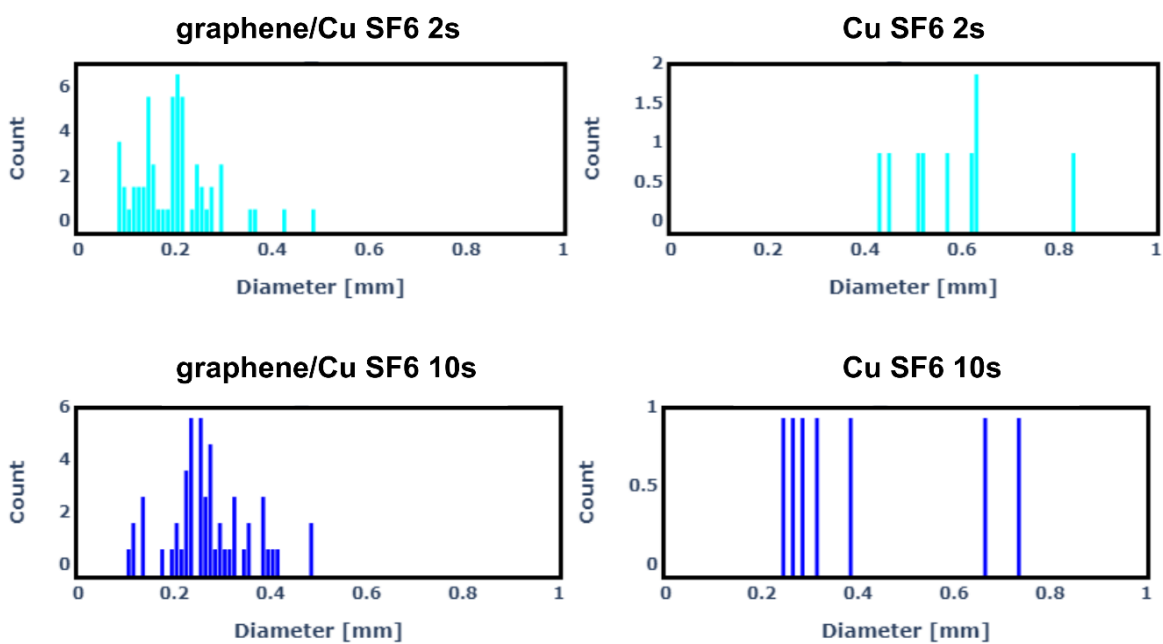

**Supporting Figure S12.** (a) Histograms of water droplet diameters for SF$_6$-plasma treated (2s top panels and 10s bottom panels) graphene/Cu (left panels) and bare annealed Cu (right panels). Showing very few and large diameter droplets for bare Cu SF$_6$-plasma treated samples.

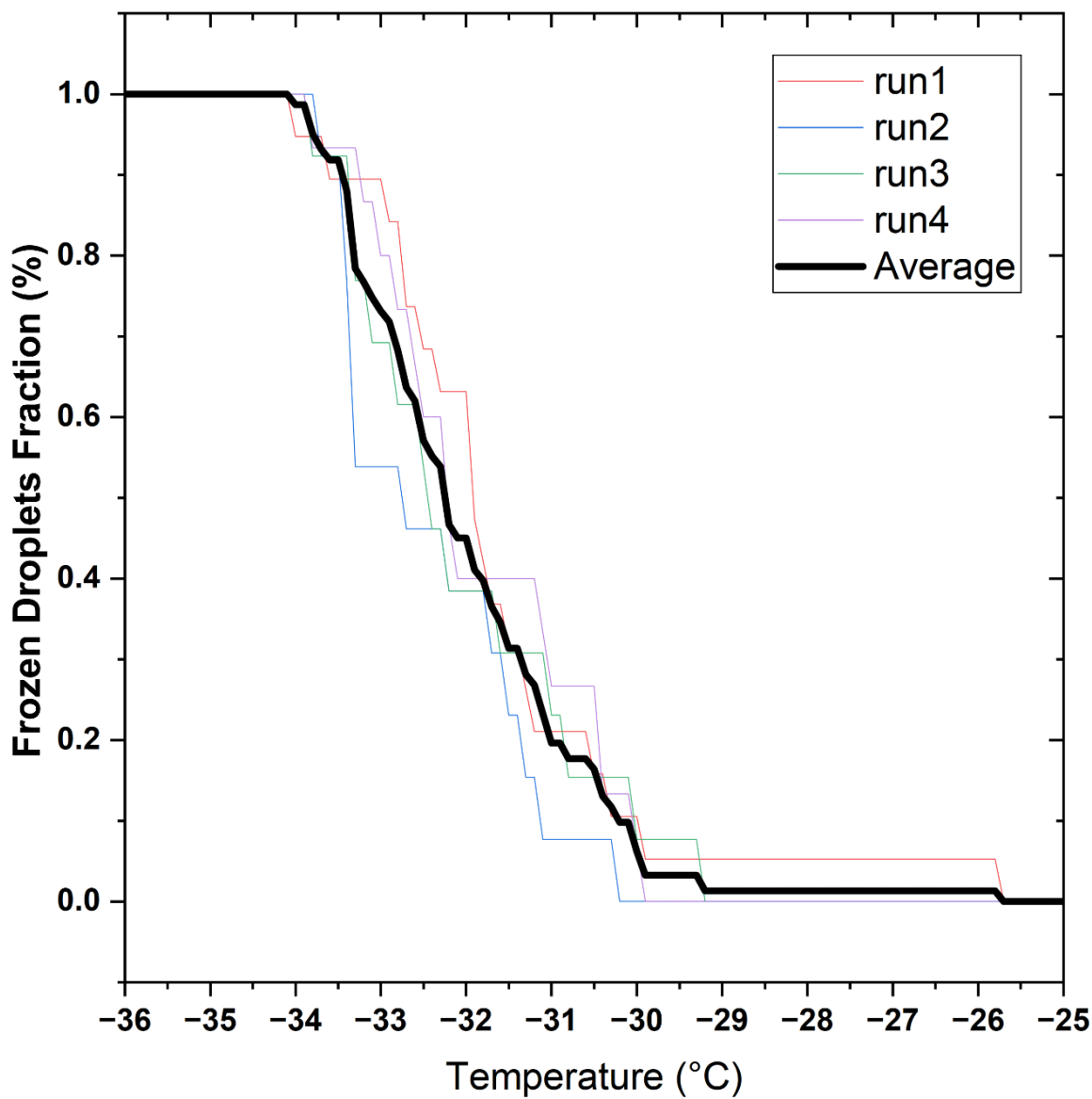

**Supporting Figure S13.** Visual depiction of our method of freezing curve average calculation exemplary for the dataset for graphene/Cu SF$_6$ plasma treated samples. For all four freezing curves (thin, faint curves), we average the frozen fraction value for each temperature value, resulting in our freezing curve average (thick, black curve). Individual freezing curves are calculated by interpolating between single droplet freezing events with a temperature step of 0.1 K. The freezing fraction is considered constant between freezing events in order to more accurately depict the actual measured datapoints, which leads to the stepwise increase of the freezing fraction with decreasing temperature.

**Supporting Table S1.** $T_{10}$, $T_{25}$ $T_{50}$ and $T_{75}$ values and their standard deviations ($\Delta T_{xx}$) for graphene/Cu and bare annealed Cu samples after various storage times in ambient.

|  | $T_{10}$/°C | $\Delta T_{10}$/°C | $T_{25}$/°C | $\Delta T_{25}$/°C | $T_{50}$/°C | $\Delta T_{50}$/°C | $T_{75}$/°C | $\Delta T_{75}$/°C |
|---|---|---|---|---|---|---|---|---|
| Graphene/Cu as-grown | –30.1 | ± 2.0 | –31.6 | ± 1.1 | –32.4 | ± 0.8 | –33.0 | ± 0.7 |
| Graphene/Cu 2 h | –25.6 | ± 2.6 | –28.1 | ± 1.7 | –30.6 | ± 1.0 | –31.8 | ± 1.2 |
| Graphene/Cu 24 h | –29.8 | ± 0.6 | –31.0 | ± 0.4 | –31.8 | ± 0.6 | –32.6 | ± 0.6 |
| Graphene/Cu 1 week | –30.1 | ± 1.1 | –31.0 | ± 0.7 | –31.6 | ± 0.5 | –32.2 | ± 0.3 |
| Graphene/Cu 1 month | –28.9 | ± 0.6 | –29.6 | ± 0.5 | –30.6 | ± 0.5 | –31.3 | ± 0.5 |
| Cu annealed | –30.5 | ± 1.4 | –31.6 | ± 1.1 | –32.3 | ± 0.9 | –32.9 | ± 0.5 |
| Cu 2 h | –23.4 | ± 5.5 | –28.1 | ± 1.3 | –29.6 | ± 0.8 | –30.4 | ± 0.7 |
| Cu 24 h | –28.9 | ± 3.7 | –31.4 | ± 1.2 | –32.4 | ± 0.6 | –32.9 | ± 0.3 |
| Cu 1 week | –30.7 | ± 0.6 | –31.9 | ± 0.9 | –32.4 | ± 0.7 | –33.1 | ± 0.5 |
| Cu 1 month | –29.2 | ± 1.8 | –31.2 | ± 1.5 | –32.2 | ± 1.5 | –32.9 | ± 0.7 |

**Supporting References**